\documentclass[journal]{IEEEtran}
\usepackage{cite}

\ifCLASSINFOpdf
  \usepackage[pdftex]{graphicx}
\else
\fi
\begin{document}
%
\title{GitHub open source project recommendation system}
%
%
%

\author{\IEEEauthorblockN{Tadej Matek\IEEEauthorrefmark{1} and
Svit Timej Zebec\IEEEauthorrefmark{2}}\\
\IEEEauthorblockA{Faculty of Computer and Information Science,
University of Ljubljana\\
Ljubljana, Slovenia\\
Email: \IEEEauthorrefmark{1}matektadej@gmail.com,
\IEEEauthorrefmark{2}svit.zebec@gmail.com}}

\maketitle

\begin{abstract}
Hosting platforms for software projects can form collaborative social networks and a prime example of this is GitHub which is arguably the most popular platform of this kind. An open source project recommendation system could be a major feature for a platform like GitHub, enabling its users to find relevant projects in a fast and simple manner. We perform network analysis on a constructed graph based on GitHub data and present a recommendation system that uses link prediction.
\end{abstract}


%
\IEEEpeerreviewmaketitle

\section{Introduction}
%
%
%
%
\IEEEPARstart{G}{itHub} \cite{github} is not only the biggest and most popular hosting platform for software projects, but also the one with the most social features. It surpasses many other alternative platforms by not only providing version control but also providing a way for developers to collaborate on open source software projects. Users can create their own public projects, to which other users can contribute, contribute to other projects, follow other users etc. In order for a user to become a contributor to another project he first has to \textit{fork} the project repository which makes his own copy of the repository that allows independent work on the project. Then he can commit changes to his own forked repository and after that contribute to the original project repository via a \textit{pull request}, which can be accepted or refused by the project owner or a project collaborator\footnote{A GitHub user role at a project level that can make changes to the project repository and review code contributions.}. If the contribution is accepted, the contributing user can be acknowledged as a contributor to the project. As a result of this GitHub forms a collaborative social network that can be interesting to analyze. 

Active users obviously contribute more, while less active users contribute less. Active users contribute to more projects than the less active users. Both GitHub user types, the active and the less active, could benefit from a system that would recommend relevant projects for them to work on. The active user is probably constantly looking for new interesting challenges and maybe even business opportunities, while the less active user might be looking to grow his collaboration network or just to find a relevant project to work on, without too much effort. The question that arises is how can one recommend a relevant project to a user using network analysis.

Our solution is to first construct a bipartite graph, whose nodes are comprised of two disjoint sets, where the nodes in the first set are GitHub projects and the nodes in the second set are GitHub users. Due to the graph being bipartite links can occur only between a node representing a project and a node representing a user, i.e.\ two nodes representing two projects or two users can not be linked. In the constructed graph, a link between a project and a user exists if the user is a contributor to the project, i.e.\ he has contributed to the project via a pull request or a commit if he is the project owner or a collaborator. We then use link prediction methods, in order to assess whether a project is relevant to the user and is ultimately worth recommending.

Next, we present the related work regarding GitHub social network, recommendation systems and link prediction for unipartite and bipartite graphs. Then, we summarize the methods used for link prediction and assessment of the quality of link prediction. We use different link prediction methods on the constructed graph representing the GitHub collaborative social network, measure their performance and recommend the best one. Finally, we discuss the results and present the final conclusions.

\section{Related work}
\label{related}

Ferdian et al. \cite{thung2013network} analyzed two sub networks of the GitHub network: a project to project network, where a project is linked to another project if it has any common developers, and a developer collaboration network. The authors report that both networks exhibit low average path lengths, with project-project network being seemingly scale-free \cite{thung2013network}. Furthermore, the study reported that the reason for low average path lengths is due to developers automatically connecting to other developers without actually knowing them.

A similar study performed developer collaboration analysis on a similar open source network \cite{surian2010mining}, namely SourceForge. They obtained similar results, the average distance appears to be very low. They also extracted topological patterns \cite{surian2010mining}, sub graphs of the original network. Through analysis of these patterns they discovered that the clustering coefficient is relatively high, i.e. the friends of a developer are likely to connect. Together with the low average distance, the SourceForge network appears to be small-world; however, we cannot infer the same for the GitHub network.

Antonio et al. \cite{lima2014coding} performed structural analysis of several GitHub networks ranging from followers network, collaboration network to geographically separated networks. They confirm our assumption that GitHub collaboration graph is also small-world, as they report a high clustering coefficient. Along with high clustering, only a fraction of projects were reported to have high amount of collaborations. This is intuitively true, as GitHub is not used merely for collaboration and social aspects, but also as a hub for storing source code and for self-promotion, so we have a lot of repositories with only one collaborator. The authors also discovered that users who are geographically close are more likely to collaborate together on a project.

The problem of recommendation has been around for quite some time. A popular approach in this area is Collaborative Filtering \cite{su2009survey} which relies on the assumption, that if two users perform similar actions, then they are also more likely to perform other actions in the same way. Several such approaches use machine learning methods, however, we are interested in methods using network structure. Recommendation systems in network analysis are usually based on link prediction techniques. 

There are several novel approaches for link prediction \cite{xie2015link, allali2011link, chinta2014cs224w, Sherkat201580}. Latest study \cite{xie2015link} uses complex numbers as edge weights, to model dual meaning of relationships in a bipartite graph. A user can either like or dislike something and authors expressed transitive relationships using complex numbers. Another recent study \cite{Sherkat201580} simulates ant colonies and their food search process. The ants randomly explore paths from their source node and upon encountering food, test it's quality. When returning to the source node, ants release pheromone. The amount of released pheromone is proportional to the quality of found food. Such approach is used to determine special sub graphs. Afterwards the evolution of found sub graphs is studied to predict links. 

Traditional link prediction approaches \cite{lu2011link} either use node similarity, which is further divided into local (common neighbors, Jaccard index, Adamic-Adar index \cite{adamic2003friends}, Resource allocation index \cite{zhou2009predicting}) and global (Katz index \cite{katz1953new}, Leicht-Holme-Newman index \cite{leicht2006vertex}, Matrix Forest index \cite{chebotarev2006matrix}), or they use maximum likelihood methods \cite{guimera2009missing}. However, most measures were adopted only for unipartite graphs. 

There have been attempts to generalize link prediction for bipartite graphs as recommendation systems usually work on such class of graphs. One study \cite{allali2011link} defines internal links, i.e. links which do not alter the amount of neighbors on the same side of bipartite graph, as a way of prediction, and report good results. Furthermore, they use weight functions to improve prediction of internal links. Weight functions can be represented by traditional node similarity measures or by additional meta information from the graph itself. A similar paper \cite{chinta2014cs224w} has extended the notion of node similarity to bipartite graphs by defining two new sets of nodes. Given two nodes $A$, $B$ on the opposite side of graph, the first set is a set of nodes at distance 2 away from node $A$, yielding all neighbors on the same side of node $A$ in the bipartite graph. The second set of nodes are nodes one hop away from node $B$, once again yielding neighbors of node $A$. The authors report quite high prediction accuracy using just simple node similarity measures.

\section{Methodology}
\label{method}

Our graph was constructed from publicly available GitHub events on GHTorrent website \cite{ghtorrent}. Because of the enormous amount of data available, we sample the original data set. We select the top 1000 repositories and all users who contributed to these top repositories. Contributions include commits and pull requests. We only consider original repositories i.e. repositories which are not forked from another repository and are still active (not deleted). The primary criterion for ranking top 1000 repositories is the amount of commits by distinct users. The more commits a repository has from different users, the higher the ranking. The resulting graph is bipartite with two distinct sets of nodes. Set $U$ represents the users which work on different projects. Set $R$ represents the top $1000$ repositories. Set of links $E$ contains all links which are present between these two distinct sets of nodes, or equivalently $E \subseteq U \times R$. All links are undirected and indicate, that a user has worked on a repository. For simplicity we keep only the largest connected component of the graph, which contains 98\% of all nodes. The resulting network contains $n = 72088$ nodes and $m = 91385$ links.

\subsection{Neighborhood representation}
\label{neighborhood_representation}

We consider two types of neighborhood representations. Because the local similarity measures \cite{lu2011link} are based on the assumption that the underlying graph is unipartite, adjustments are necessary for bipartite setting. The first representation is based on the redefinition of the neighborhood of nodes in one of two distinct sets \cite{chinta2014cs224w}. Let $u \in U$ denote a user node and $r \in R$ a repository node. We redefine the traditional neighborhood $\Gamma(u)$ of user node to a new neighborhood $\Gamma_{new}(u)$, according to Eq. \ref{eq:nghbr}. It is clear from the equation that we look at all the nodes two steps away from the user node. In addition, we redefine the degree of a user node as $k'_u = \left\vert{\Gamma_{new}(u)}\right\vert$.

\begin{equation}
\label{eq:nghbr}
\Gamma_{new}(u) = z; z \in \Gamma(r) \wedge r \in \Gamma(u) \wedge z \neq u
\end{equation}

We keep the traditional neighborhood of repository node $\Gamma(r)$ as all nodes one step away. Such definition allows us to find intersections between user and repository nodes, when predicting a link. Local similarity measures once again apply and are used to predict the likelihood of user and repository connecting.

Next we consider a more advanced generalization of link prediction in bipartite graphs. Specifically, we use the notion of internal links \cite{allali2011link}. First we consider a bottom projection $G_{b}$ of our graph $G$, which consists of user nodes only. A link is present among two user nodes in $G_{b}$, if both nodes share common neighbors in the original graph $G$. Internal link is defined as a link $l \in E$, if it does not change the bottom projection $G_b$ \cite{allali2011link}. Each internal link induces several links in the $G_b$. Furthermore, we can specify weights for all links in bottom projection, $E_b$, using arbitrary weight function. Only internal links are predicted. An internal link is predicted when and if it induces any link in $G_b$, which has a weight larger than a predefined threshold. The basic assumption of the approach is that a user node and repository node are more likely to connect in the future, if they have lots of common neighbors. The converse is also true; user and repository are unlikely to connect if they do not share common neighbors.

The weight functions can also be represented by local similarity measures. No redefinition of neighborhood is necessary, as all comparisons are based on two user nodes. In addition, meta information can be used to define the similarity of two users.

\subsection{Prediction evaluation}
\label{prediction_evaluation}

Link prediction can be seen as unsupervised learning. Given a subset of links $E_{pos} \subset E$ (positive examples) and a set of links $E_{neg}$ not present in $E$ (negative examples), the goal of any link prediction method is to predict as many links from $E_{pos}$ (called true positives) and as few links from $E_{neg}$ (called true negatives). Two additional errors are present, the false positives, where a method predicts the link to exist while the link is in fact in $E_{neg}$ and conversely the false negatives. A standard measure of performance is AUC (Area Under the ROC Curve). ROC (Receiver Operating Characteristic) curve depicts a ratio between true positive rate (also known as sensitivity or recall) and false positive rate (also known as the fall-out). The larger the AUC, the better the ratio of true positive rate versus false positive rate.

The following framework is used to calculate AUC. We first randomly sample $\frac{\left\vert{E}\right\vert}{10}$ pairs of nodes $U \times R$ which are not yet linked. These are used as negative examples. Next we randomly sample and remove $\frac{\left\vert{E}\right\vert}{10}$ links from the original graph $G$. These are used as positive examples. We compute, for each node pair in positive and negative examples, the similarity index, according to a specific measure. AUC is then defined as the probability, that a randomly chosen positive example has a higher similarity index than a randomly selected negative example. We randomly sample $\frac{\left\vert{E}\right\vert}{10}$ pairs from positive and negative examples (with repetitions) and compare their similarity indexes. Let $m_1$ be the number of times the similarity index of a positive example is higher than the index of a negative example and let $m_2$ denote the number of times the indices are equal. AUC is then calculated according to Eq. \ref{eq:auc}.

\begin{equation}
\label{eq:auc}
AUC = \frac{m_1 + \frac{m_2}{2}}{\frac{\left\vert{E}\right\vert}{10}}
\end{equation}

The worst AUC is $0.5$, which means that the classifier's answers are completely random. AUC lower than $0.5$ can be reverted to high AUC by inverting the answers of the classifier.

\subsection{Local similarity measures}
\label{local_similarity_measures}

All local similarity measures are based on neighborhoods of two nodes. We consider the similarity indices in Table \ref{simIndex}. The neighborhood $\Gamma$ and degree of node $k$ in equations in Table \ref{simIndex} depend on the type of neighborhood definition in bipartite graphs. The Common neighbors index is defined as the size of common neighborhood of two nodes. Jaccard index is very similar to Common neighbors index but takes into account also the total size of neighborhoods of both nodes. Hub Promoted index \cite{ravasz2002hierarchical} promotes hub nodes, i.e. the nodes with high degree, while the Hub Depressed index \cite{ravasz2002hierarchical} does the opposite. Motivated by the scale-free property, where a node is likely to connect to another node with probability proportional to the degree of the node, the Preferential Attachment index is derived from the exact same property. The Adamic-Adar index \cite{adamic2003friends} also counts the common neighbors, but assigns each common neighbor a weight according to connectedness \cite{su2009survey}. Resource Allocation index \cite{zhou2009predicting} is motivated by the flow of resources through the network of common neighbors \cite{su2009survey}, where each node receives a fraction of resources from another node through their common neighbors. Salton Cosine index \cite{chowdhury2010introduction} measures the cosine similarity of two nodes (or vectors).

We also consider weighted combination of local similarity measures. The assumption is, that local measures themselves are unable to accurately capture the variety of structural properties of the network. A combination of local measures includes several structural properties at a time. Furthermore, if a classifier based on local similarity measures produces AUC smaller than $0.5$, we invert it's answers by taking $-s(x, y)$.

\begin{table}[!h]
\renewcommand{\arraystretch}{2.0}
\caption{Local similarity measures for link prediction}
\label{simIndex}
\centering
\begin{tabular}{ll}
\hline
\textbf{Measure} & \textbf{Definition} \\
\hline
Common neighbors index & $s(x, y) = \left\vert{\Gamma(x) \cap \Gamma(y)}\right\vert$\\
Jaccard index & $s(x, y) = \frac{\left\vert{\Gamma(x) \cap \Gamma(y)}\right\vert}{\left\vert{\Gamma(x) \cup \Gamma(y)}\right\vert}$\\
Hub Promoted index \cite{ravasz2002hierarchical} & $s(x, y) = \frac{\left\vert{\Gamma(x) \cap \Gamma(y)}\right\vert}{min(k_x, k_y)}$\\
Hub Depressed index \cite{ravasz2002hierarchical} & $s(x, y) = \frac{\left\vert{\Gamma(x) \cap \Gamma(y)}\right\vert}{max(k_x, k_y)}$\\
Preferential Attachment index & $s(x, y) = k_x \times k_y$\\
Adamic-Adar index \cite{adamic2003friends} & $s(x, y) = \sum_{z \in \Gamma(x) \cap \Gamma(y)} \frac{1}{log k_z}$\\
Resource Allocation index \cite{zhou2009predicting} & $s(x, y) = \sum_{z \in \Gamma(x) \cap \Gamma(y)} \frac{1}{k_z}$\\
Salton Cosine index \cite{chowdhury2010introduction} & $s(x, y) = \frac{\left\vert{\Gamma(x) \cap \Gamma(y)}\right\vert}{\sqrt{k_x \times k_y}}$\\
\hline
\end{tabular}
\end{table}

\subsection{Communities}
\label{communities}

We use the Infomap algorithm \cite{rosvall2008maps} to detect communities in our graph. Afterwards we use another measure, similar to local similarity measures, which uses community information to predict the likelihood of two users connecting. Given two nodes, similarity is defined according to Eq. \ref{eq:comm}, where $m_C$ represents the number of links in community $C$, $M_C$ the maximum possible number of links within the community, $\frac{m_C}{M_C}$ the link density in community and $\delta$ the standard Kronecker's delta.

\begin{equation}
\label{eq:comm}
s(x, y) = \frac{m_C}{M_C} \times \delta_{c_x, c_y}
\end{equation}

\section{Results}
\label{results}

\begin{figure}[!t]
\centering
\includegraphics[width=0.45\textwidth]{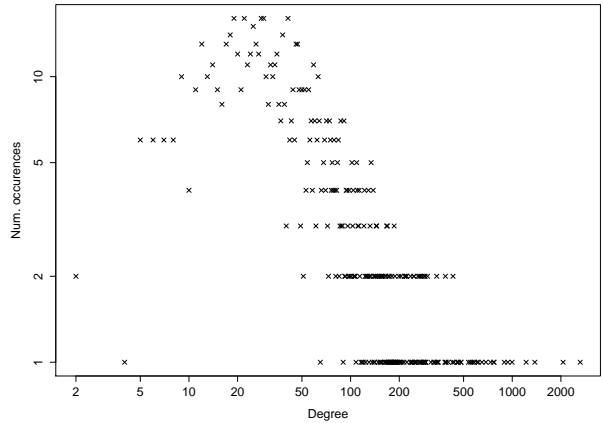}
\caption{Log-log plot of degree distribution of the repository nodes in the largest connected component.}
\label{repos}
\end{figure}

\begin{figure}[!t]
\centering
\includegraphics[width=0.45\textwidth]{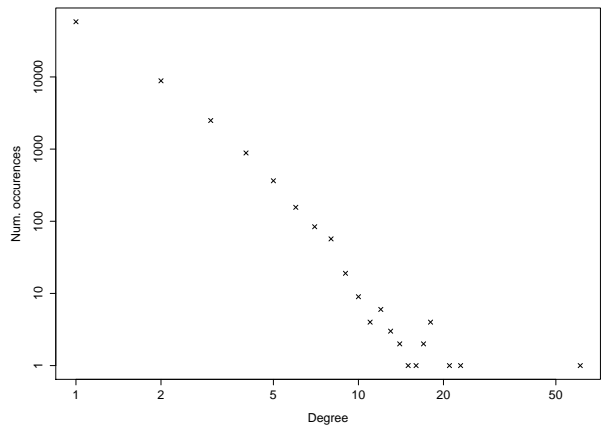}
\caption{Log-log plot of degree distribution of the user nodes in the largest connected component.}
\label{users}
\end{figure}

According to Figure \ref{users}, the network is seemingly scale-free only for user nodes, however we cannot be certain from the graph alone. The average degree of user nodes, $2.8$, indicates that there are a lot more users/collaborators than repositories and because users on average work only on two projects, the average degree is low. On the other hand, repositories are mostly hubs with many users working on them. However, in spite of the fact that repositories are hubs, they show no signs of power-law property, according to Figure \ref{repos}. The average clustering coefficient is, as expected, equal to zero, because there are no closed triangles in the network.

\subsection{Extended neighborhood for user nodes}
\label{ExtNeighForUsers}

The user node neighborhood is represented as all the nodes two steps away as described in Section \ref{neighborhood_representation}. This applies to all measures in this subsection. Out of the local measures, the best one turned out to be the Preferential Attachment index with inverted answers, yielding an AUC of $0.72$. Inverting answers is a technique described in Section \ref{local_similarity_measures}. Other local measures either yielded AUC $< 0.5$, similar to Preferential Attachment index, or AUC near $0.60$.

In order to try to boost the performance of the link prediction we combined different local measures as described in Section \ref{local_similarity_measures}. After trying different combinations of local measures and weights we found that the combination of Preferential Attachment index and Adamic-Adar index, where the former has a weight of $0.7$ and the latter a weight of $0.3$, yields a high AUC of $0.89$. The definition of this combined index can be seen in Eq. \ref{eq:combined}, where $s(x, y)_{PA}$ is the Preferential Attachment index and $s(x, y)_{AA}$ is the Adamic-Adar index. 
\begin{equation}
\label{eq:combined}
s(x, y)_{comb} = 0.7 \times s(x, y)_{PA} + 0.3 \times s(x, y)_{AA}
\end{equation}

Finally we used a measure that utilizes community information and is described in Section \ref{communities}. This measure yielded an AUC of $0.90$.

A summary of all AUC results for link prediction measures using the extender neighborhood representation for user nodes can be observed in the Table \ref{tab:AUCResults}.

\begin{table}[!h]
\renewcommand{\arraystretch}{2.0}
\caption{AUC results for link prediction measures using the extended neighborhood representation for user nodes as described in Section \ref{neighborhood_representation}}
\label{tab:AUCResults}
\centering
\begin{tabular}{ll}
\hline
\textbf{Measure} & \textbf{AUC} \\
\hline
Common neighbors index & $0.59$\\
Jaccard index & $0.60$\\
Hub Promoted index \cite{ravasz2002hierarchical} & $0.27$\\
Hub Depressed index \cite{ravasz2002hierarchical} & $0.59$\\
Preferential Attachment index & $0.27$\\
Preferential Attachment index (inverted) & $0.72$\\
Adamic-Adar index \cite{adamic2003friends} & $0.60$\\
Resource Allocation index \cite{zhou2009predicting} & $0.60$\\
Salton Cosine index \cite{chowdhury2010introduction} & $0.28$\\
Community measure using Infomap \cite{rosvall2008maps} & $\textbf{0.90}$\\
Combined measure: $70\%$ Preferential Attachment \\ index and $30\%$ Adamic-Adar index & $0.89$\\

\hline
\end{tabular}
\end{table}

\subsection{Internal link prediction}

All the local similarity measures used as a weight function with internal link prediction yield AUC of $0.5$. The problem however, is not in the local similarity measures but the internal links themselves. There are, on average, only $15$ internal links predicted compared to $9000$ links which the method should predict, even with threshold set to zero. We discuss the problem in the next section.

\section{Discussion}

It appears that none of the local similarity measures is able to capture the network's structure, because the highest obtained AUC is equal to $0.72$ and the others are significantly worse. This means that these local measures are not suitable to be used in a recommendation system, because the AUC is too low. We are able to capture the network's structure in a better way with a weighted combination of local measures though. This can be observed in Section \ref{ExtNeighForUsers} where Preferential Attachment index and Adamic-Adar index are combined and the combined measure yields an AUC of $0.89$. Given all that, it should be noted, that this applies only to the sampled network. As a result of that, this combined measure is also not suitable to be used in a recommendation system, because the original non-sampled GitHub network can be different and this measure may not be able to capture  the network's structure in such a good way.

The AUC for community based link prediction turns out to be the best. It appears that strong communities form in our sampled GitHub network. Further investigation reveals, that approximately $50$ communities form in the largest connected component, as detected by the Infomap algorithm. The assumption, that if two users are part of the same community, they are likely to be interested in the same projects, is not so unlikely, given the high prediction accuracy. Considering that open source projects are community-oriented, this is not a surprise. Behind every large project, there is a community of users which collaborate and share code. Moreover, the users from such communities are also more likely to work together on other similar projects. The communities revealed in our network can also be viewed as groups of similar interests. This is also why local measures perform poorly - their effect is local and cannot capture interests of a larger group of users. The community based link prediction is suitable to be used in a recommendation system.

We anticipated the internal links to give an alternative representation of node similarity in our bipartite graph and therefore help with the prediction process. However, it turns out that the low amount of predicted internal links can be explained by our graph's structure. Because our network is a source code collaboration network and given the degree distributions in Figures \ref{users} and \ref{repos}, it is clear why this is the case. An average developer works only on two projects. The consequence of this is that the amount of common neighbors with other users is extremely low. Since internal link framework relies on the assumption that two nodes are more likely to connect if they have a lot of common neighbors, our graph is not suited for such method. Nonetheless the method itself is interesting and could be potentially used on other types of networks with high connectivity, such as an actors network.

\section{Conclusion}
In this paper we analyze different link prediction methods on a graph, that is sampled from a GitHub network, in order to determine which method is the best and can be used for a GitHub open source project recommendation system. To evaluate the performance of link prediction methods we use AUC, as described in Section \ref{prediction_evaluation}.

We use two different approaches to represent the neighborhood. These two are an extended neighborhood, where the neighborhood of the user node are all nodes two steps away and an internal link representation, as described in Section \ref{neighborhood_representation}. We evaluate the local similarity measures, using both neighborhood representations. Using the extended neighborhood representation, most measures yield an AUC of $0.60$ and the best one yields an AUC of $0.72$. Using the internal link representation, all measures yield an AUC of $0.50$, which determines that the internal link representation is useless in our case. We achieve an AUC of $0.89$ with combining multiple local measures, but we disregard this method, because it performs this well only on this specific graph.

The best method turns out to be community based link prediction, as described in Section \ref{communities}, with an AUC of $0.90$. This link prediction method has very high performance, due to the formation of strong communities in the GitHub network and can be used in a system for recommending open source projects to users on GitHub.


%




\ifCLASSOPTIONcaptionsoff
  \newpage
\fi



\bibliographystyle{IEEEtran}
%
\bibliography{references}

%





\end{document}